# Analytical Models For Galactic Nuclei

HongSheng Zhao

hsz@mpa-garching.mpg.de

Max-Planck-Institut für Astrophysik,

Karl-Schwarzschild-Straße 1,

85740 Garching-bei-München, Germany

## ABSTRACT

I present a general family of dynamical models with simple analytical potential-density pairs suited to model galactic bulges and nuclei with double power-law radial density profiles and an optional central black hole. Analytical expressions for the potential and velocity dispersion of these models are given in the spherical case. Many previously known analytical spherical models, including also the recent $\gamma/\eta$-models by Dehnen (1993) and Tremaine et al. (1994), are special cases of this family. This family also forms a complete set for constructing general galaxy models or solving Poisson's equation in the non-spherical case. In particular, a generalized Clutton-Brock (1973) and Hernquist-Ostriker (1992) orthogonal basis set is given.

*Subject headings:* celestial mechanics, stellar dynamics -galaxies: kinematics and dynamics - galaxies: nuclei, methods: analytical

## 1. Introduction

Recent high resolution photometric data from the Hubble Space Telescope (HST) (Crane and Stiavelli 1993, Kormendy et al. 1994, Ferrarese et al. 1994) reveal three basic features in the mass distribution of galactic nuclei: (1) a central cusp in radial density profile, $\rho \propto r^{-\gamma}$ with $0 < \gamma < 3$; (2) a markedly non-spherical shape (flattened/triaxial/lopsided) in density distribution (except for perhaps the nearly spherical M87 nucleus); (3) a possible massive central black hole, which is suggested by the cusped light profile (Young 1976). Of these systems, M32 and M31 are two examples. The former has a flattened E3 nucleus with a central cusp $\gamma \sim 1.5$. The latter has a double-nucleus, where the fainter peak (P2) appears to be cusped in the high resolution HST observation and the off-centered brighter peak (P1) may generate a significant lopsided perturbation (m=1 mode) in the potential (Lauer et al. 1993). For both a massive central black hole of roughly a few times $10^6 M_\odot$ and $10^7 M_\odot$ respectively has been proposed based on in detailed axisymmetric models of kinematic data (Dehnen 1994, Qian et al. 1995); the results are in qualitative agreement with earlier spherical models (Dressler and Richstone 1988).

Modelling the structure of nearby galactic nuclei can help us constrain their formation history and understand their relation with distant active galactic nuclei. A new class of models are needed to account for the above three basic features in recent observations. The simple parametrization first proposed by Hernquist (1990) appears to be an attractive choice. It is a dimensionless spherical volume density model with three free parameters $(\alpha, \beta, \gamma)$,

$$\rho(r) = \frac{C}{r^\gamma (1 + r^{1/\alpha})^{(\beta-\gamma)\alpha}}, \qquad (1)$$

where $C$ is a normalization constant. This model parametrizes the volume density as a general double power-law with slope $-\gamma$ at radii much smaller or $-\beta$ at radii much bigger than the break radius $r = 1$. The third parameter, $\alpha$, is a measure of the width of the transition region: the bigger $\alpha$ is, the wider the transition region. Similar parametrizations have been used to fit observations of galactic nuclei (Kormendy et al. 1994, Crane and Stiavelli 1993).

Hernquist (1990) proposed the $(\alpha, \beta, \gamma)$-models only as a density model. Their dynamical properties are not known analytically except for the Hernquist bulge and a few other well-known special cases (see Table I), and, more recently, a narrow sequence due to Dehnen (1993) and Tremaine et al. (1994). The latter authors found that the spherical $(1, 4, \gamma)$ subset has many good analytical properties, including analytical potential, intrinsic velocity dispersion for all real values of $\gamma$ between 0 and 3, and for $\gamma = 0, 1, 2$ analytical projected light and dispersion as well, all of which remain analytical when a central black hole is introduced. I shall later on call this $(1, 4, \gamma)$ subset as $\gamma/\eta$-models, which relate to the $\eta$-models by Tremaine et al. (1994) by $\gamma \equiv 3 - \eta$. Note throughout the paper, $\gamma$ always denotes the inner power-law slope in the volume density, $\beta$ the outer power-law slope. All models have a finite mass. The total luminous mass and the gravitational constant $G$ are always set to unity.

In this paper, I show that the potential of the whole set of $(\alpha, \beta, \gamma)$ density models can be computed analytically, and the isotropic velocity dispersion of most of these spherical models with or without a black hole can be given analytically. Table I shows that the family includes many well-known spherical models, e.g., the Hernquist bulge (Hernquist 1990), the Jaffe model (Jaffe 1983), the Plummer model (1911), the modified Hubble profile (Binney and Tremaine 1987), the modified isothermal dark halo model (Sackett and Sparke 1990), the perfect sphere model (de Zeeuw 1985a,b) and the recent $\gamma/\eta$ series (Dehnen 1993,

Tremaine et al. 1994). But most of the models in the three parameter spherical family are previously unknown. Some have very simple potentials, like the $\alpha$ model and the $\beta$ model. The $(n, 3 + \frac{k}{n}, \gamma)$ three paramemter family also have analytical velocity dispersions. Besides a series of models included in the known $\gamma/\eta$ set, there are also two new models with analytical distribution functions.

The simple analytical properties of the spherical models make them useful for i) deriving the mass-to-light ratio (Tremaine et al. 1994) as traditionally done by core fitting with a King model, ii) setting up equilibrium conditions for N-body simulations to examine the secular evolution of galactic nuclei (Quinlan et al. 1995) and iii) testing numerical models (Dehnen and Gerhard 1994). While modelling the light and kinematic data of galactic nuclei is generally a numerical problem, analytical potentials can still serve as a solver for Poisson's equation and speed up the force calculation in N-body simulations (Hernquist and Ostriker 1992, Hernquist et al. 1995) and in building orbit models (Zhao 1994).

Section 2 shows that the potential of the general $(\alpha, \beta, \gamma)$ density models can generally be written analytically for both the spherical and the non-spherical cases. Section 3 presents the exact analytical expressions and asymptotic expressions for the dynamical quantities in the new spherical models. A brief summary is in Section 4. Many less important details of the models are in the Appendix.

## 2. Analytical potentials of $(\alpha, \beta, \gamma)$-models

Table I lists various new analytical models and known models which are subsets of the spherical $(\alpha, \beta, \gamma)$-models. One can see that the $\gamma/\eta$-models belong to a three paramamter $(n, 3 + \frac{k}{n}, \gamma)$ family which has analytical dispersions; $n$ and $k$ are any natural numbers. The latter is one of the four spherical families whose potentials can be expressed in terms of elementary functions. The potential can always be reduced to the incomplete Beta-function for both the spherical and the non-spherical case if the radial profile of each angular term in the density model is a double power-law with generally different slopes.

## 2.1. Two simple examples

Before showing the details of the general models, I give two simple examples. It is straightforward to apply the Poisson's equation to the following two potentials,

$$\Phi(r) = -\frac{1}{r}(1 - \frac{1}{(1+r)^{\beta-3}}), \qquad (2)$$

and

$$\Phi(r) = -\frac{1}{(1+r^{1/\alpha})^\alpha}, \qquad (3)$$

and to show that their corresponding densities have double power-law profiles. The former makes up a new analytical potential-density pair which I call the $(1, \beta, 1)$-models, or in short the $\beta$-models. These models have a fixed inner density profile of $r^{-1}$ but a variable outer slope $\beta$ in contrast to the $\gamma/\eta$ models, which have a fixed outer profile $r^{-4}$ but a variable inner profile. The latter analytical potential-density pair is the $(\alpha, 3+\frac{1}{\alpha}, 2-\frac{1}{\alpha})$ models, which I call the $\alpha$-models. They fill the gap between the familiar Plummer model (Plummer 1911; Binney & Tremaine 1987) with $\alpha = \frac{1}{2}$ and the Hernquist model with $\alpha = 1$.

The limitations of the $\alpha$-models, $\beta$-models and the $\gamma/\eta$-models are obvious when it comes to fitting observations. Generally one cannot expect a good fit for all three regions of a spherical nucleus, namely, a central cusp, a power-law tail and the transition region, with only one free parameter. The central bulge and nucleus region of our Galaxy, for example, has a double power-law of slope $-1.8$ and $-3.7$ inside and outside 1 kpc respectively (Sellwood and Sanders 1988), which cannot be well-approximated by any of these simple models. And like the central regions of M32 and M31, the system is far from being spherical based on recent near infrared maps of the Galactic plane from *COBE* (de Zeeuw 1993, Dwek et al. 1995). For these reasons, I will focus on the more interesting subsets of the spherical $(\alpha, \beta, \gamma)$-models which have simple analytical expressions for dynamical quantities but flexible density profiles. The generalization to the non-spherical case is straightforward.

## 2.2. General analytical spherical subsets

Consider the potential for the spherical $(\alpha, \beta, \gamma)$ density models. Substituting equation 1 to equation (2-122) of Binney and Tremaine (1987), one finds that

$$\Phi(r) = -4\pi C(\int_r^\infty r^{-\gamma-1}(1+r^{1/\alpha})^{-(\beta-\gamma)\alpha}dr + r^{-1}\int_0^r r^{2-\gamma}(1+r^{1/\alpha})^{-(\beta-\gamma)\alpha}dr). \qquad (4)$$

After a change of variable from $r$ to $\chi$,

$$\chi \equiv \frac{1+\xi}{2} = \frac{r^{1/\alpha}}{r^{1/\alpha}+1}, \tag{5}$$

the integrals in the expression for the potential can be written in terms of the incomplete Beta-function,

$$\Phi(r) = -4\pi\rho(r)f_{0,0}(r)r^2, \tag{6}$$

$$f_{0,0}(r) = \frac{\alpha B(c_0-q_0, q_0, \chi)}{\chi^{c_0-q_0}(1-\chi)^{q_0}} + \frac{\alpha B(c_0-p_0, p_0, 1-\chi)}{(1-\chi)^{c_0-p_0}\chi^{p_0}} \tag{7}$$

where the constants $p_0$, $q_0$ and $c_0$ are certain combinations of $\alpha$, $\beta$ and $\gamma$ given in the Appendix. From the properties of the incomplete Beta-function (see the Appendix), one can prove that *the model potential $\Phi$ reduces further to elementary functions if ONE of the following four pairs of numbers is a pair of natural numbers:*
$$(c_0-p_0, q_0), (p_0, c_0-q_0), (p_0, q_0) \text{ and } (c_0-p_0, c_0-q_0).$$

Among the four spherical families, the $\gamma/\eta$-models, $\beta$-models and $\alpha$-models shown earlier belong to the first three families with the natural number pair being $(2,1)$, $(1,2)$ and $(1,1)$ respectively. Other explicit expressions will be given only for the family with $(c_0-p_0, q_0)$ being a pair of natural numbers $(n+k, k)$, which is the most interesting family because it has analytical velocity dispersion as well as potential. See equation 17 of Section 3.1..

### 2.3. Generalization in shape and radial profile

The analytical expression for the potential is also readily generalizable to the non-spherical case. This requires that the density of the observed system in polar coordinates (r, $\theta$, $\phi$) can be decomposed into spherical harmonics $Y_{l,m}(\theta, \psi)$

$$\rho(r, \theta, \psi) = \sum_{l,m} \sqrt{4\pi} Y_{l,m}(\theta, \psi) \rho_{l,m}(r), \tag{8}$$

and each angular component (l,m) has a double power-law radial profile

$$\rho_{l,m}(r) = \frac{C_{l,m}}{r^{\gamma_l}(1+r^{1/\alpha_l})^{(\beta_l-\gamma_l)\alpha_l}}, \tag{9}$$

where $(\alpha_l, \beta_l, \gamma_l)$ are three parameters to be adjusted to fit observation at the transition, inner and outer regions respectively. The corresponding potential has an analytical expression very similar to that of the spherical model,

$$\Phi(r,\theta,\psi) = -\sum_{l,m} \frac{4\pi}{2l+1}\sqrt{4\pi} Y_{l,m}(\theta,\psi)\rho_{l,m}(r)f_{l,m}(r)r^2, \tag{10}$$

$$f_{l,m}(r) = \frac{\alpha_l B(c_l-q_l,q_l,\chi)}{\chi^{c_l-q_l}(1-\chi)^{q_l}} + \frac{\alpha_l B(c_l-p_l,p_l,1-\chi)}{(1-\chi)^{c_l-p_l}\chi^{p_l}}. \tag{11}$$

In the spherical case, equation 8, 10 and 11 become equation 1, 6 and 7, as the coefficients $C_{l,m}$ are all zero except for a normalization constant $C_{0,0} = C$, which is given in the Appendix. The subscript $l$ or $(l,m)$ is sometimes omitted in the spherical models, where only the $(l,m) = (0,0)$ term is relavent. To simplify notation, a subscript $m$ is also omitted for the parameters $p_l$, $q_l$ and $c_l$, which can depend on the azimuthal number $m$. The potential again reduces to elementary functions if one of the following is a pair of natural numbers

$$(c_l - p_l,\ q_l),\ (p_l,\ c_l - q_l),\ (p_l,\ q_l) \text{ and } (c_l - p_l,\ c_l - q_l).$$

For an observed system whose the density profile can not be well-fitted by a single double power-law (equation 1), it is still of numerical interests to fit the profile with a few double power-law components because the total potential will then be written as a linear combination of several explicitly known analytical components. The computation of force can be faster than fully numerical methods, particularly if the observed system has a double power-law profile to the zeroth order. The same is true for the angular expansion of the density distribution of a general non-spherical nucleus. In fact the whole family of $(\alpha,\beta,\gamma)$-models and their spherical harmonics terms form a complete basis set for solving the Poisson's equation. The orthogonal basis functions shown next are linear combinations of the subset of the $(\alpha,\beta,\gamma)$-models with $(p_l,q_l)$ being natural numbers. Although the whole $(\alpha,\beta,\gamma)$ set is degenerate and non-orthogonal, it can still be useful in new methods to solve Poisson's equation with a general non-orthogonal basis set (Saha 1993). However, I will focus on the orthogonal set, which is easier to apply.

### 2.4. $\alpha$-model and orthogonal basis functions for Poisson's equation

An efficient Poisson solver is crucial in massive N-body simulations since the accelerations of particles need to be computed at each time step. One of the most efficient methods is the Self-Consistent Field (SCF) method by Hernquist and Ostriker (1992). In this method, the potential and the density are written as a superposition of orthogonal basis functions, which are solutions of the Poisson's equation. The gain in speed depends on carefully choosing the lowest order term, so that one is able to resolve the density and the potential in the least number of expansion terms. The lowest order term in the Clutton-Brock (1973) expansion is the Plummer model, which is suitable for systems with a finite core. Hernquist and Ostriker (1992) generalize the method so that the lowest order term can be the Hernquist model, which well approximates the $R^{\frac{1}{4}}$ law distribution of ellipticals and bulges. Since galactic nuclei neither have cores nor follow a unique $R^{\frac{1}{4}}$ law, these two known expansions are clearly not always the optimal.

One can easily build an optimal basis set by choosing the lowest order term to be the more general $\alpha$-model shown in the Section 2.1.. Depending on whether the interesting region is the central cusp or the halo, one can choose the value of $\alpha$ according to the inner power-law slope $\gamma = (2 - \frac{1}{\alpha})$ or the outer slope $\beta = (3 + \frac{1}{\alpha})$. The orthogonal basis functions are given as follows in notations similar to Hernquist and Ostriker (1992).

$$\rho_{n,l,m}(r,\theta,\psi) = \frac{\tilde{K}_{n,l}\sqrt{4\pi}Y_{l,m}(\theta,\psi)}{r^{2-1/\alpha}(1+r^{1/\alpha})^{2+\alpha}} \frac{r^l}{(1+r^{1/\alpha})^{2l\alpha}} G_n^w(\xi), \qquad (12)$$

$$\Phi_{n,l,m}(r,\theta,\psi) = -\frac{\sqrt{4\pi}Y_{l,m}(\theta,\psi)}{(1+r^{1/\alpha})^{\alpha}} \frac{r^l}{(1+r^{1/\alpha})^{2l\alpha}} G_n^w(\xi), \qquad (13)$$

where $G_n^w(\xi)$ is the Gegenbauer polynomials, $\xi$ is given in equation 5 and the constants $\tilde{K}_{n,l}$, $N_{n,l}$ and $w$ are given in the Appendix. One recovers the Hernquist-Ostriker expansion and the Clutton-Brock expansion by letting $\alpha = 1$ and $\frac{1}{2}$ respectively. Such expansions have been applied to approximate the potential of the boxy Galactic bar (Zhao 1994) based on the *COBE* observation (Dwek et al. 1995). The application of such models in the fully parallel Self-Consistent Field code (Hernquist et al. 1995) remains to be seen.

## 3. Dynamics of analytical spherical $(\alpha, \beta, \gamma)$-models

A new family of dynamical models that are analytical in both velocity dispersion and potential with or without a central point mass can be obtained within the spherical $(\alpha, \beta, \gamma)$-models. This property allows us to study the dynamics of a cusped nucleus under the influence of a central black hole analytically under the assumption that the phase space density depends only on the energy integral so that the velocity dispersion $\sigma^2(r)$ of the system is everywhere isotropic. Much of the calculation can be done in the same fashion as in Tremaine et al. (1994).

### 3.1. The analytical $(n, 3 + \frac{k}{n}, \gamma)$ subset with/without a black hole

I find that in many cases $\sigma(r)$, $\Phi(r)$, $M(r)$ and $\rho(r)$ can all be written as elementary functions of $r$. The most general and useful subset of these analytical models is the $(n, 3 + \frac{k}{n}, \gamma)$-models with an optional central point mass, where $n$ and $k$ are any natural numbers and $\gamma$ is a real number between 0 and 3. Also any linear combination of these models (with the same value of $\alpha$) is also analytical. These models are less restrictive than the $(1, 4, \gamma)$ models, and greatly broaden the range of analytical self-consistent models with black holes. The projected density and dispersion are also analytical if one restricts oneself to the $(1, 3 + k, 0)$, $(1, 3 + k, 1)$ and $(1, 3 + k, 2)$-models. See Table 1.

For the $(n, 3 + \frac{k}{n}, \gamma)$-models with a constant mass-light ratio $\Upsilon$ and a central black hole of mass $m_{BH}$,

$$\rho(r) = C\chi^{-\alpha\gamma}(1-\chi)^{\alpha\beta}, \tag{14}$$

$$M(r) = m_{BH} + 4\pi\alpha C B(\alpha(3-\gamma), \alpha(\beta-3), \chi), \tag{15}$$

$$= m_{BH} + \sum_{i=0}^{\alpha(\beta-3)-1} a_i \chi^{\alpha(3-\gamma)+i}, \tag{16}$$

$$\Phi(r) = -\frac{m_{BH}}{r} - \sum_{i=0}^{\alpha(\beta-2)-2} b_i S_{\alpha(2-\gamma)+i}(\chi), \tag{17}$$

$$\sigma^2(r) = \frac{\alpha\chi^{\alpha\gamma}}{(1-\chi)^{\alpha\beta}} \int_\chi^1 d\chi_1 M(r_1(\chi_1))\chi_1^{-\alpha(1+\gamma)-1}(1-\chi_1)^{\alpha(\beta+1)-1}, \tag{18}$$

$$= \frac{1}{\rho(r)}(\sum_{i=0}^{2\alpha(\beta-1)-2} d_i S_{2\alpha(1-\gamma)+i}(\chi) + m_{BH} \sum_{i=0}^{\alpha(\beta+1)-1} e_i S_{\alpha(-1-\gamma)+i}(\chi)), \tag{19}$$

$$I(R) = \frac{1}{\Upsilon R} \sum_{i=0}^{\alpha(2-\gamma)} f_i U^{(\alpha)}_{\alpha(\beta-2)+i}(R), \tag{20}$$

$$\sigma_p^2(R) = \frac{1}{\Upsilon I(R)}(\sum_{i=0}^{\alpha(2\beta-3)-1} g_i V^{(\alpha)}_{\alpha(3-2\gamma)+i}(R) + m_{BH} \sum_{i=0}^{\alpha\beta} h_i V^{(\alpha)}_{-\alpha\gamma+i}(R)), \tag{21}$$

where the coefficients $a_i$, $b_i$, $d_i - h_i$ and the functions $S_i(\chi)$, $U_i^{(\alpha)}$ and $V_i^{(\alpha)}$ are given in the Appendix. $\chi$ is given in equation 5. The above formulae can be derived by changing variable from $r$ to $\chi$ followed by integrating a polynomial expansion with respect to $\chi$. For the $(1, 4, \gamma)$-model, the above reduces to the formulae in Tremaine et al. (1994).

There are many simple and plausible families in the analytical $(n, 3 + \frac{k}{n}, \gamma)$-models, for example, the $(1, 4, \gamma)$-model, $(1, 5, \gamma)$-model, $(1, 6, \gamma)$-model, $(2, \frac{7}{2}, \gamma)$-model, $(2, \frac{9}{2}, \gamma)$-model, $(2, \frac{11}{2}, \gamma)$-model. As an illustration, I give the closed form of these analytical properties for the $(1, 5, \gamma)$ models, which have a steeper outer power-law than the $\gamma/\eta$ model.

$$\rho(r) = \frac{(\gamma - 3)(\gamma - 4)}{4\pi r^\gamma (1 + r)^{5-\gamma}}, \tag{22}$$

$$M(r) = m_{BH} + y^{3-\gamma}((4 - \gamma) - (3 - \gamma)y), \tag{23}$$

$$\Phi(r) = -\frac{m_{BH}}{r} - (4 - \gamma)S_{2-\gamma}(y) + (3 - \gamma)S_{3-\gamma}(y), \tag{24}$$

$$\sigma^2(r) = \frac{y^\gamma}{(1-y)^5}[(4 - \gamma)S_{2-2\gamma}(y) - (23 - 6\gamma)S_{3-2\gamma}(y) + 5(10 - 3\gamma)S_{4-2\gamma}(y)$$
$$- 10(7 - 2\gamma)S_{5-2\gamma}(y) + 5(10 - 3\gamma)S_{6-2\gamma}(y) - (19 - 6\gamma)S_{7-2\gamma}(y) + (3 - \gamma)S_{8-2\gamma}(y)$$
$$+ m_{BH}(S_{-1-\gamma}(y) - 5S_{-\gamma}(y) + 10S_{1-\gamma}(y) - 10S_{2-\gamma}(y) + 5S_{3-\gamma}(y) - S_{4-\gamma}(y))], \tag{25}$$

where $y = \chi = r/(1+r)$. Although the expression for $\sigma_r$ is long, it involves no cancellation of large terms and hence is stable when evaluated numerically; each $S_i(y)$ term is well-behaved mathematically and is finite except at the origin $y = r = 0$ for $i \leq 0$.

Analytical phase space densities are also available for somewhat restrictive but plausible models without a central black hole. These are related to the analytical potential-density pairs through the Eddington's formula (Binney and Tremaine 1987). Besides the Plummer model and the $(1, 4, 2 - \frac{1}{n})$ series for all natural number values of $n$ (Dehnen 1993, Tremaine et al. 1994), I find two more models with this property: the $(1, 5, 1)$-model with

$$f(\epsilon) = \frac{9}{\sqrt{2}\pi^3}[-\frac{\sqrt{\epsilon}(33 + 70\epsilon - 114\epsilon^2 + 32\epsilon^3)}{3(1 + 4\epsilon)(2 - \epsilon)^2}$$
$$+ \arctan 2\epsilon + \frac{3}{2(2 - \epsilon)^{5/2}}(\arctan \frac{3\sqrt{\epsilon}}{\sqrt{2 - \epsilon}} + \arctan \frac{\sqrt{\epsilon}}{\sqrt{2 - \epsilon}})], \tag{26}$$

and the $(2, \frac{7}{2}, \frac{3}{2})$-model with

$$f(\epsilon) = \frac{27}{64\sqrt{2}\pi^2}[-1 + \frac{\sqrt{\epsilon}(18 - 69\epsilon - 30\epsilon^2 - 40\epsilon^3 + 16\epsilon^4)}{9\pi(1 - \epsilon)^4}$$
$$+ \frac{6 - 27\epsilon + 56\epsilon^2}{6(1 - \epsilon)^{9/2}}(1 + \frac{2\arcsin \sqrt{\epsilon}}{\pi})], \tag{27}$$

where $-\epsilon$ is the energy integral. As a nucleus with an adiabaticly grown central black hole would have a central cusp of power $\frac{3}{2}$ (Binney and Tremaine 1987), the $(2, \frac{7}{2}, \frac{3}{2})$ model can be of some interest. The $(1, 5, 1)$-model is very similar to the Hernquist model and can be useful for bulges with density fall-off steeper than the $R^{1/4}$ law at large radii. For alternative approaches to building models with analytical distribution functions, see Dejonghe (1984).

It can be easily shown that all of the $(\alpha, \beta, \gamma)$-models have a positive definite distribution function $f(E) \geq 0$ based on the Eddington's formula (Binney & Tremaine 1987) since $\frac{d^2\rho}{d\Phi^2} \geq 0$. The stability of these models is unclear. A sufficient condition for stability is $\frac{df(E)}{dE} < 0$ or more directly $\frac{d^3\rho}{d\Phi^3} \leq 0$ (e.g., Binney and Tremaine 1987). The models are likely to be stable in the absence of a black hole: one can show that the sufficient condition is met for certain subsets of the models, e.g., the $\alpha$-model, $\beta$-model and $\gamma/\eta$-models; one can also show this is the case for all models at very small and very large radius. Some models with a black hole have $\frac{df(E)}{dE} > 0$ (Tremaine et al. 1994) and the stability of these models is as yet unknown.

### 3.2. Asymptotic expressions at small and large radius

Asymptotic expressions are available for all $(\alpha, \beta, \gamma)$-models. In particular, the asymptotic expression for the surface light is also a double power-law, as for the volume density, but with a slope $1 - \beta$ at large radii and $1 - \gamma$ at small radii (if $\gamma > 1$); if $0 < \gamma < 1$, the cusp only appears in the volume density. So the asymptotic behavior in the integrated light of the $(\alpha, \beta, \gamma)$-models are nearly the same as the set of double power-law models used by Kormendy (1994) to fit the observed galactic nuclei. The properties of the $(\alpha, \beta, \gamma)$-models are also qualitatively the same as the $\gamma/\eta$-model at small radii and can be classified into three types as in Tremaine et al. (1994). But the models are much more general than the $\gamma/\eta$-model at large radii, because the asymptotic volume density profile is $r^{-\beta}$ rather than the fixed $r^{-4}$, hence they can fit a wider range of observations. The asymptotic expressions for $\sigma^2(r)$, $\sigma_p^2(R)$ and $I(R)$ are given as follows. A black hole of mass $m_{BH}$ is included.

At small radii,

$$\sigma^2(r) \to A_1 r^\gamma + \frac{m_{BH}}{(1+\gamma)r} \quad \text{if } \gamma < 1, \tag{28}$$

$$\to A_{10} r \log\frac{1}{r} + \frac{m_{BH}}{(1+\gamma)r} \quad \text{if } \gamma = 1, \tag{29}$$

$$\to A_2 r^{2-\gamma} + \frac{m_{BH}}{(1+\gamma)r} \quad \text{if } \gamma > 1, \tag{30}$$

$$I(R) \to A_3 \quad \text{if } \gamma < 1, \tag{31}$$

$$\to A_{11} \log\frac{1}{R} \quad \text{if } \gamma = 1, \tag{32}$$

$$\to A_4 R^{1-\gamma} \quad \text{if } \gamma > 1, \tag{33}$$

$$\sigma_p^2(R) \to A_5 + \frac{m_{BH} A_8}{R^\gamma} \quad \text{if } \gamma < 1, \tag{34}$$

$$\to \frac{A_{12}}{\log\frac{1}{R}} + \frac{m_{BH} A_{14}}{R \log\frac{1}{R}} \quad \text{if } \gamma = 1, \tag{35}$$

$$\rightarrow \quad A_6 R^{1-\gamma} + \frac{m_{BH} A_9}{R} \quad \text{if } \frac{3}{2} > \gamma > 1, \tag{36}$$

$$\rightarrow \quad A_{13} R^{\frac{1}{2}} \log \frac{1}{R} + \frac{m_{BH} A_9}{R} \quad \text{if } \gamma = \frac{3}{2}, \tag{37}$$

$$\rightarrow \quad A_7 R^{2-\gamma} + \frac{m_{BH} A_9}{R} \quad \text{if } \gamma > \frac{3}{2}, \tag{38}$$

where the constants $A_1 - A_{14}$ are given in the Appendix.

At large radii,

$$\sigma^2(r) \rightarrow \frac{1 + m_{BH}}{(1+\beta)r}, \tag{39}$$

$$I(R) \rightarrow \frac{D_4}{R^{\beta-1}}, \tag{40}$$

$$\sigma_p^2(R) \rightarrow D_9 \frac{1 + m_{BH}}{R}, \tag{41}$$

where the constants $D_4$ and $D_9$ equal to $A_4$ and $A_9$ respectively if replacing $\gamma$ with $\beta$.

## 4. Summary

In conclusion, I have presented a general family of models with analytical potential-density pairs which have asymptotic behaviors the same as observed galactic nuclei. More specificly, the density distributions have a double power-law radial profiles, which can be used to fit the central cusp, the power-law tail and the transition regions of observed galactic nuclei. In the spherical case the models generalize the $\gamma/\eta$-models by Dehnen (1993) and Tremaine et al. (1994), and greatly enlarge the set of models with good analytical dynamical properties. Analytical expressions and asymptotic expressions for the intrinsic and projected velocity dispersion of the models are given for cases with and without a central black hole.

The structure and dynamics of a general galactic nuclei are often more complex than in a spherical isotropic double power-law model. Both the non-spherical distribution in the light and a possible anisotropy in the velocity need to be modelled numerically, e.g., with the Schwarzschild (1979, 1982) approach and with the N-body approach. However, the analytical potentials are still very useful as they can be generalized to the non-spherical case and applied in these numerical methods as a fast solver for Poisson's equation.

I thank R. Michael Rich and David N. Spergel for encouragements and comments on an earlier draft, Hans-Walter Rix, Simon White and Walter Dehnen for discussions and critical readings of the manuscript, John Hibbard for giving me the reference to the $\gamma/\eta$-models.

## 5. Appendix

In the following, I give detailed expressions for a few quantities used in the main text.

1. The three parameters $p_l$, $q_l$ and $c_l$ in equation 7 and 11.

$$p_l = \alpha_l(-\gamma_l - l + 2) \,, q_l = \alpha_l(\beta_l - l - 3) \,, c_l = \alpha_l(\beta_l - \gamma_l) \,. \tag{42}$$

2. The incomplete Beta-function $B(a, b, x)$.

$$B(a, b, x) = \int_0^x dt\, t^{a-1}(1-t)^{b-1} \,. \tag{43}$$

It is a simple integration, which reduces to elementary functions of $x$ if $a$ or $b$ or $1 - a - b$ is a natural number. One can prove this for the case that $b$ is a natural number by expanding the polynomial $(1-t)^{b-1}$ and performing the integration for each term. Similarly for the other two cases after a change of variable from $t$ to $1 - t$ or $(1/t - 1)$.

In cases that the function cannot be reduced to elementary ones or the expression is too lengthy, it can be numerically computed by a function call to, e.g., the efficient BETAI program in *Numerical Recipes* (Press et al. 1992); it converges in $O(\sqrt{Max(a,b)})$ iterations.

3. The normalization constant $C$.

$$C = \frac{1}{4\pi\alpha B(\alpha(3-\gamma), \alpha(\beta-3), 1)} \,. \tag{44}$$

4. The constants $\tilde{K}_{n,l}$, $N_{n,l}$ and $w$ in equations 12 and 13.

$$\int dr^3 \rho_{n',l',m'}(r,\theta,\psi)\Phi_{n,l,m}(r,\theta,\psi) = \delta_{n,n'}\delta_{l,l'}\delta_{m,m'} N_{n,l}, \tag{45}$$

$$\tilde{K}_{n,l} = \frac{4(n+w)^2 - 1}{16\pi\alpha^2}, \tag{46}$$

$$\frac{1}{N_{n,l}} = -\frac{2^{4w+1}\alpha(n+w)}{\pi(4(n+w)^2 - 1)} \frac{n!\Gamma^2(w)}{\Gamma(2w+n)}, \tag{47}$$

where

$$w = (2l+1)\alpha + \frac{1}{2} \,. \tag{48}$$

5. The constants $A_1 - A_{14}$ in equations 28 to 38.

$$A_1 = 4\pi\alpha^2 C \int_0^1 x^{-\alpha(1+\gamma)-1}(1-x)^{\alpha(\beta+1)-1}dx \int_0^x x_1^{\alpha(3-\gamma)-1}(1-x_1)^{\alpha(\beta-3)-1}dx_1$$

$$A_2 = \frac{2\pi C}{(3-\gamma)(\gamma-1)},$$

$$A_3 = 2\alpha C B(\alpha(1-\gamma), \alpha(\beta-1), 1), \quad A_4 = CB(\frac{\gamma-1}{2}, \frac{1}{2}, 1),$$

$$A_5 = \frac{4\pi\alpha C \int_0^1 x^{-\alpha\gamma-1}(1-x)^{\alpha\beta-1}dx \int_0^x x_1^{\alpha(3-\gamma)-1}(1-x_1)^{\alpha(\beta-3)-1}dx_1}{B(\alpha(1-\gamma), \alpha(\beta-1), 1)},$$

$$A_6 = \frac{A_5 A_3}{A_4}, \quad A_7 = \frac{4\pi\alpha C B(\gamma-\frac{3}{2}, \frac{3}{2}, 1)}{(3-\gamma)B(\frac{\gamma-1}{2}, \frac{1}{2}, 1)}, \quad (49)$$

$$A_8 = \frac{A_9 A_4}{A_3}, \quad A_9 = \frac{B(\frac{\gamma}{2}, \frac{3}{2}, 1)}{B(\frac{\gamma-1}{2}, \frac{1}{2}, 1)}, \quad A_{10} = \frac{1}{2B(2\alpha, \alpha(\beta-3), 1)},$$

$$A_{11} = \frac{A_{10}}{\pi\alpha}, \quad A_{12} = \frac{A_3 A_5}{A_{11}}, \quad A_{13} = \frac{8A_{10}}{3B(\frac{1}{4}, \frac{1}{2}, 1)}, \quad A_{14} = \frac{A_9 A_4}{A_{11}}$$

6. $S_i(\chi)$ in equations 14 to 21.

$$S_i(\chi) = -\log\chi \quad \text{if } i = 0 \quad (50)$$
$$= \frac{1-\chi^i}{i} \quad \text{otherwise.} \quad (51)$$

where $\chi$ is given in equation 5. The function is singular at $\chi = 0$ for $i \leq 0$.

7. The coefficients $a_i$, $b_i$, $d_i - h_i$ in equations 14 to 21.

$$a_i = \frac{4\pi\alpha C}{\alpha(3-\gamma)+i}q(\alpha(\beta-3)-1, i), \quad b_i = \alpha\sum_{j=0}^{i} q(\alpha-1, j)a_{i-j}, \quad d_i = \sum_{j=0}^{i} e_j a_{i-j},$$

$$e_i = \alpha C q(\alpha(\beta+1)-1, i), \quad f_i = 2Cq(\alpha(2-\gamma), i), \quad g_i = \sum_{j=0}^{i} h_j a_{i-j}, \quad h_i = 2Cq(\alpha\beta, i),$$

$$q(i,j) = (-1)^j \frac{i!}{j!(i-j)!}, \quad \text{if } i \geq j \geq 0 \text{ , otherwise} = 0. \quad (52)$$

8. The functions $U_i^{(\alpha)}(R)$ and $V_i^{(\alpha)}(R)$ in equations 20 and 21.

$$U_i^{(\alpha)}(R) = \int_0^{\pi/2} \left(\frac{\sin^{1/\alpha} u}{R^{1/\alpha} + \sin^{1/\alpha} u}\right)^i du, \quad (53)$$

$$V_i^{(\alpha)}(R) = \int_0^{\pi/2} \left(\frac{R^{1/\alpha}}{R^{1/\alpha} + \sin^{1/\alpha} v}\right)^i \cos^2 v \, dv. \quad (54)$$

One can further reduce $U_i^{(\alpha)}(R)$ and $V_i^{(\alpha)}(R)$ to elementary functions for all values of $i$ by the following recursion relation for the $\alpha = 1$ sequence.

$$U_{i+1}^{(1)}(R) = U_i^{(1)}(R) + \frac{R}{i}\frac{dU_i^{(1)}(R)}{dR}, \tag{55}$$

$$V_{i+1}^{(1)}(R) = V_i^{(1)}(R) - \frac{R}{i}\frac{dV_i^{(1)}(R)}{dR}, \tag{56}$$

and

$$U_1^{(1)}(R) = -RZ(R) + (\frac{\pi}{2} - 1)R, \tag{57}$$

$$V_1^{(1)}(R) = -(1 - R^2)Z(R) + (\frac{\pi}{2} + 1)R^2 - R - 1, \tag{58}$$

$$V_0^{(1)}(R) = \frac{\pi}{4}, \tag{59}$$

$$V_{-1}^{(1)}(R) = \frac{1}{3R} + \frac{\pi}{4}, \tag{60}$$

$$V_{-2}^{(1)}(R) = \frac{2}{3R} + \frac{\pi}{16R^2} + \frac{\pi}{4}, \tag{61}$$

where

$$Z(R) = RX(R) - 1 \tag{62}$$

$$X(R) = 1 \text{ if } R = 1, \tag{63}$$

$$= \frac{\cosh^{-1}(1/R)}{(1 - R^2)^{1/2}} \text{ if } R < 1, \tag{64}$$

$$= \frac{\arccos(1/R)}{(R^2 - 1)^{1/2}} \text{ otherwise}. \tag{65}$$

Note that near $R = 1$, stable numerical evaluation of $X(R)$ needs to make use of the the Taylor expansion given in Hernquist (1990).

Table 1: Analytical models in the $(\alpha, \beta, \gamma)$-family

| Analytical quantities | parameters $(\alpha, \beta, \gamma)$ | Comments |
|---|---|---|
| $\rho(r, \theta, \psi)$ | $(\alpha, \beta, \gamma)^*$ | $\rho(r) = \frac{C}{r^\gamma (1 + r^{1/\alpha})^{(\beta-\gamma)\alpha}}$ |
| | | non-spherical: eq. 8 and 9 |
| $\Phi(r, \theta, \psi)$ | $(n, 3 + \frac{k}{n}, \gamma)$ | eq. 17 |
| | and others† | non-spherical: eq. 10 and 11 |
| $\sigma^2(r)$ and $M(r)$ | $(n, 3 + \frac{k}{n}, \gamma)$ | eq. 19 and 16 with optional BH |
| $\sigma_p^2(R)$ and $I(R)$ | $(1, 3+k, 0)$ | eq. 20 and 21 with optional BH |
| | $(1, 3+k, 1)$ | |
| | $(1, 3+k, 2)$ | |
| $f(E)$ | $\gamma/\eta$-models with $\gamma = 2 - \frac{1}{n}$ | Dehnen 1993/Tremaine et al. 1994 |
| | $(\frac{1}{2}, 5, 0)$ | Plummer (1911) |
| | $(1, 5, 1)$ | eq. 26 without BH |
| | $(2, \frac{7}{2}, \frac{3}{2})$ | eq. 27 without BH |
| Special cases: | | |
| $\alpha$-models | $(\alpha, 3 + \frac{1}{\alpha}, 2 - \frac{1}{\alpha})$ | $\Phi(r) = -\frac{1}{(1+r^{1/\alpha})^\alpha}$ |
| $\beta$-models | $(1, \beta, 1)$ | $\Phi(r) = -\frac{1}{r}(1 - \frac{1}{(1+r)^{\beta-3}})$ |
| $\gamma/\eta$-models | $(1, 4, \gamma)$ | $\Phi(r) = -\frac{1 - (1+r^{-1})^{\gamma-2}}{2-\gamma}$ |
| Hernquist model (1990) | $(1, 4, 1)$ | cusped |
| Jaffe model (1983) | $(1, 4, 2)$ | cusped |
| Plummer model (1911) | $(\frac{1}{2}, 5, 0)$ | finite core |
| Perfect sphere | $(\frac{1}{2}, 4, 0)$ | finite core (de Zeeuw 1985a,b) |
| Mod. Hubble profile | $(\frac{1}{2}, 3, 0)$ | finite core (Binney & Tremaine 1987) |
| Mod. isothermal sphere | $(\frac{1}{2}, 2, 0)$ | finite core (Sackett and Sparke 1990) |

$^*$ $\alpha$, $\beta$ and $\gamma$ are the transition width at the break radius $r = 1$, the slope of the outer and inner power-law respectively. Also $G = M = 1$.

† If $n$ and $k$ are any natural numbers, 1, 2, 3, etc, or if one of the four pairs $(c_l - p_l, q_l)$, $(p_l, c_l - q_l)$, $(p_l, q_l)$ and $(c_l - p_l, c_l - q_l)$ is a pair of natural numbers; $p_l$, $q_l$ and $c_l$ are defined in eq. 42 and $l$ is the angular quantum number.